\documentclass[12pt]{article}
 \newcommand{\be}[1]{\begin{equation}\label{#1}}
 \newcommand{\ba}[1]{\begin{eqnarray}\label{#1}}

 \newcommand{\rd}{{\rm d}}
 
 \newcommand{\re}{{\rm e}}
 \newcommand{\pa}[1]{\left(#1\right)}
 \newcommand{\paq}[1]{\left[#1\right]}

 \newcommand{\M}{{\rm M_{\rm P}}}

 \def\ee{\end{equation}}
 \def\ea{\end{eqnarray}}

\usepackage{amsmath}

\usepackage{graphicx,latexsym}
\usepackage{graphicx,epsf}
\usepackage{color}
\usepackage{epsfig}
\usepackage{amsmath}
\usepackage[T1]{fontenc}
\usepackage[utf8]{inputenc}
\usepackage{tikz}
\usepackage{authblk}
\begin{document}
\title{Scale Invariant Dark Energy}
\author[1]{Alessandro Tronconi\thanks{alessandro.tronconi@bo.infn.it}}
\author[1]{Giovanni Venturi\thanks{giovanni.venturi@bo.infn.it}}
\affil[1]{Dipartimento di Fisica e Astronomia and INFN, Via Irnerio 46, 40126 Bologna, Italy}

\date{}

\maketitle

\begin{abstract}
A global scale-invariant Dark Energy model based on Induced Gravity with the addition of a small $R^2$ contribution is examined. The scalar field (quintessence), playing the role of Dark Energy, has a quartic potential and generates Newton's constant with its non-minimal coupling (after introducing a suitable symmetry breaking). Even when small, the $R^2$ contribution significantly modifies the cosmological evolution of the matter-gravity system. The solutions to this model are obtained analytically through a perturbative expansion and oscillate with transplanckian frequency. They are then compared with similar solutions found for $\Lambda$CDM cosmology plus $R^2$. Finally scalar field production is perturbatively taken into account in a simple model and the resulting effects illustrated.
\end{abstract}
\section{Introduction}
Alexei Starobinsky was a brilliant and original theoretical physicist who made fundamental contributions to gravitation and cosmology. He visited our group in Bologna for the first time in 2003 and left a lasting scientific influence on us. This manuscript is dedicated to his memory.\\

Even if gravity is an attractive force, at least for the short distances we can directly probe, the current understanding of theoretical cosmology is limited by the fact that several independent observations suggest that our universe underwent, and is currently going through, a phase of cosmic expansion and acceleration. The acceleration is the consequence of a ``repulsive'' and long range effect which perhaps originates from some modification of General Relativity or the presence of some exotic ``invisible'' fluid. The inflationary phase is the period of accelerated expansion which occurred before the Hot Big Bang, and dynamically produced the initial conditions of the present Universe. While the exact microscopic physics at the origin of inflation is still debated, its dynamics seems well constrained by observations and we generally expect that, in the near future as such observations become more precise, we shall gain more insights on inflation and the origin of our universe. In contrast there is much uncertainty about the physics behind the present cosmic acceleration. While a cosmological constant or a mechanism similar to inflation may explain, to a certain degree, current observations, several questions related to such an accelerated evolution remain unresolved, above all the tension between the differing measured values of the present rate of expansion of the universe ($H_0$). This is certainly the reason behind the abundant theoretical and experimental efforts which try to improve the present understanding of the current expansion of the universe.\\
Several years ago a simple model based on global scale invariance was suggested for
induced gravity \cite{Cooper,Turchetti}. The model involved a massless scalar field coupled to the curvature scalar and having a quartic self interaction. The introduction of symmetry breaking, either through the presence of a condensate \cite{Zee} or quantum effects \cite{Sakharov}, would then lead to a non-zero expectation value for the scalar field. As a consequence one would then generate a gravitational constant and a de Sitter solution associated with a cosmological constant \cite{Kam}. The addition of an
isentropic perfect fluid as a perturbation would then lead to a time dependence of the
gravitational constant which, because of the stabilizing influence of the potential, is much
smaller than that (power-law) of previous models (Brans-Dicke) \cite{Brans}. Some years ago, since the
observational status was compatible with an accelerating universe dominated by dark
energy, we studied the cosmological evolution of the model in the presence of matter and
radiation. We found that it led to Einstein gravity plus a cosmological constant as a
stable attractor among homogeneous cosmologies and was therefore a viable dark energy
model for a wide range of scalar field initial conditions and values for its positive coupling to
the Ricci scalar \cite{FTV,Ballardini}. Subsequently the same model was also studied in the context of inflation \cite{Cerioni} with a symmetry breaking potential for the scalar field, necessary to provide the exit from the accelerated inflationary phase.\\
Several years ago de Sitter solutions were also found from the homogeneous Einstein
equations with quantum one-loop contributions of conformally covariant matter fields.
This essentially corresponded to the addition of an $R^2$, scale invariant, contribution to the Einstein action obtaining the so-called Starobinsky model\cite{Staro}. The Starobinsky model has been studied thoroughly as a model of inflation and essentially is essentially indistinguishable from induced gravity with a Higgs-like potential. Modified gravity models inspired by the seminal work \cite{Staro} was also proposed in order to explain the present universe acceleration. In such $f(R)$ models, Dark Energy is a consequence of the modified gravity dynamics and in general, in order to have acceleration for ``small'' values of the curvature scalar, one needs rather exotic actions. Moreover such models has been found to have problems which can be cured by the addition of an $R^2$ contribution \cite{DeFelice}\\
It is then natural to ask oneself what the effect is on late time cosmological evolution of adding such a scale invariant term to the previous dark energy action \cite{Cooper}: this is the purpose of this note. \\
We shall then compare some features of the cosmological evolution of a globally scale-invariant model, with $R^2$ contribution added, to the evolution one finds in standard $R+R^2/m^2$ theory with a cosmological constant (we shall refer to this latter case as perturbed $\Lambda$CDM model). In particular two effects are studied: we shall first consider the classical solutions to the matter gravity system and compare such solutions with $R+R^2/m^2$ cosmology; subsequently we shall account for possible scalar particle production effects and again compare the scale invariant model with $R+R^2/m^2$. Lastly we shall study the asymptotic behaviour of the homogeneous solutions of the two models. We end by summarizing the differences between the two models and our conclusions. Let us add that the scale invariant model under consideration, has already been studied in the context of inflation \cite{Rinaldi}. As an inflationary model it belongs to the class of multi-field inflation models since the addition of $R^2$ introduces one more scalar degree of freedom. 
\section{$R+R^2$ Modified Gravity Formalism}
We first consider standard cosmological fluids in an expanding universe with the following gravitational action 
\be{grS}
S=\frac{\M^2}{2}\int \rd^4x\sqrt{-g}\pa{R+\frac{\alpha R^2}{2\M^2}}\equiv \frac{\M^2}{2}\int \rd^4x\sqrt{-g}f(R)
\ee
where $R$ is the Ricci scalar. Modified gravity models, wherein the Einstein-Hilbert action is replaced by some generic function of the Ricci scalar, have been studied thoroughly in the literature in order to describe inflation and Dark Energy. Here we shall only consider the effects of the scale invariant contribution $\sim R^2$ on the $\Lambda$CDM cosmological evolution. If one considers a spatially flat FRW space-time the metric tensor is defined by
\be{metric}
\rd s^2={g_{\mu\nu}\rd x^\mu\rd x^\nu}=-\rd t^2+a(t)^2\rd \vec x\cdot \rd \vec x
\ee
and $t$ is the cosmic time. For such a homogeneous and isotropic metric the Ricci scalar takes the form $R=6\pa{2H^2+\dot H}$. The variation of the matter-gravity action leads to the following set of equations
\be{fr00ii}
3FH^2=\frac{FR-f}{2}-3 H\dot F+\frac{\rho}{\M^2},\quad -2F\dot H=\ddot F-H\dot F+\frac{\rho+P}{\M^2}
\ee
where $F=\partial f/\partial R$ and $\rho$ and $P$ are the total energy density and pressure of the cosmological fluids, each satisfying the continuity equation
\be{Teq}
\dot \rho_i=-3 H(\rho_i+P_i)=-3H\rho_i(1+w_i)
\ee
and the index $i=r,m,\Lambda$ stands for radiation, matter and cosmological constant.\\
One can combine the Eqs. (\ref{fr00ii}) to obtain the following equation
\be{scalGR}
\ddot R+3 H \dot R+\frac{\M^2}{3\alpha}R=\frac{\rho-3P}{3\alpha}
\ee
having the form of the Klein Gordon equation for the Ricci scalar with a mass term and an external force term depending on trace of the energy momentum tensor of the cosmological fluids. 
In this case, the $00$ component of the Einstein Eqs. (\ref{fr00ii}) is sufficient to determine the cosmological evolution. In terms of the Hubble parameter squared it has the form
\be{frH2}
0=1-\frac{\rho}{3\M^2H^2}+\frac{3\alpha}{\M^2}\paq{(H^2)_{,NN}-\frac{\paq{(H^2)_{,N}}^2}{4H^2}+3(H^2)_{,N}}
\ee
where $N\equiv \ln a$ is the e-folding number. The Friedmann equation in the context of $f(R)$ theories of gravity is a dynamical equation and its $\alpha\rightarrow 0$ limit is non-trivial. Therefore for $\alpha\ll 1$ we expect that cosmological solution can differ significantly to the standard ($\alpha=0$) one: $3\M^2H^2=\rho$.
If we search for solutions which are close to the $\Lambda$CDM evolution then one can start from the ansatz
\be{perth2}
H^2=\frac{\rho}{3\M^2}(1+\delta h_2)
\ee 
and obtain the following linearised equation for $\delta h_2$, valid in the regime $\delta h_2\ll 1$:
\ba{dh2}
0&=&\frac{(3\rho_m+4\rho_r)^2}{\rho^2}\rho^2\delta h_{2,\rho\rho}+\paq{\frac{3(3\rho_m+4\rho_r)^2}{2\rho^2}+4\frac{\rho_r}{\rho}}\rho\,\delta h_{2,\rho}\nonumber\\
&+&\paq{-\frac{(3\rho_m+4\rho_r)^2}{2\rho^2}+8\frac{\rho_r}{\rho}+\frac{\M^4}{\alpha\rho}}\delta h_{2}
+4\frac{\rho_r}{\rho}-\frac{(3\rho_m+4\rho_r)^2}{4\rho^2}
\ea
where $\rho=\rho_r+\rho_m+\rho_\Lambda$.
\subsection{Radiation domination}
During the radiation domination epoch $\rho_m/\rho\ll 1$, $\rho_r/\rho\simeq 1$ and Eq. (\ref{dh2}) simplifies and takes the form of an equation for a time dependent harmonic oscillator
\be{dh2R2}
\delta h_2''+\frac{\delta h_2}{9 x^{2/3}}=0
\ee
where the prime denotes a derivative with respect to $x=\pa{\frac{\M^4}{\alpha \rho}}^{3/4}$ and, for realistic models where the total energy density in many order of magnitude less than the Planck energy density $\M^4$, one must consider the regime $x\gg 1$.
The general solution of (\ref{dh2R2}) is
\ba{solR}
\!\!\!\delta h_2(x)&=&\sqrt{x}\pa{A_1 J_{-\frac{3}{4}}\pa{\frac{x^{\frac{2}{3}}}{2}}+A_2 J_{\frac{3}{4}}\pa{\frac{x^{\frac{2}{3}}}{2}}}\simeq \frac{x^{2/3}}{2}\pa{B_1 \sin\frac{x^{2/3}}{2}+B_2 \cos\frac{x^{2/3}}{2}}\nonumber\\
&=&\frac{\M^2}{2\sqrt{\alpha \rho}}\pa{B_1 \sin\frac{\M^2}{2\sqrt{\alpha \rho}}+B_2 \cos\frac{\M^2}{2\sqrt{\alpha \rho}}}.
\ea
Apart from the oscillation we observe that the amplitude of the solution increases with time. The perturbative approach then fails unless the integration constants are set to tiny values and we may refer to this as fine tuning. Such a fine tuning problem can be alleviated if one takes into account the effect of perturbative particle production (see for example \cite{Dolgov}). In such a case the energy density of the oscillating field is transferred to matter fields and the amplitude of the oscillation is therefore damped.\\
\subsection{Matter domination}
During matter domination $\rho_r/\rho\ll 1$, $\rho_m/\rho\simeq 1$ and Eq. (\ref{dh2}) becomes
\be{dh2M2}
\delta h_2''+\pa{\frac{4}{9}-\frac{2}{x^2}}\delta h_2-\frac{1}{x^2}=0
\ee
which can be solved exactly and in the large $x$ limit has the following general solution
\be{solM}
\delta h_2(x)\simeq A_1 \cos\frac{2x}{3}+A_2 \sin\frac{2x}{3}=A_1 \cos\frac{2\M^2}{3\sqrt{\alpha \rho}}+A_2 \sin\frac{2\M^2}{3\sqrt{\alpha \rho}}
\ee
which is an oscillation with constant amplitude. If $\delta h_2\ll 1$ at the onset of matter domination, it remains small and the perturbative solution is valid until dark energy begins to dominate. \\
Let us note that the oscillating contributions in (\ref{solR}) and (\ref{solM}) have the form
\be{osc}
\sin \frac{2\M^2}{3(1+w_{\rm eff})\sqrt{\alpha \rho}}\quad {\rm or }\quad \cos \frac{2\M^2}{3(1+w_{\rm eff})\sqrt{\alpha \rho}}
\ee
where $w_{\rm eff}$ is the equation of state of dominant fluid (and is 0 for dust and 1/3 for radiation). From the continuity equation $\dot \rho=-3 H\rho(1+w_{\rm eff})$ one obtains
\be{coneq}
\rho^{-3/2}\dot \rho=-\frac{\sqrt{3}}{\M}(1+w_{\rm eff})\Rightarrow \frac{1}{\sqrt{\rho}}=\frac{\sqrt{3}(1+w_{\rm eff})(t-t_0)}{2\M}
\ee
and on substituting into Eq. (\ref{osc}) one finds the following constant frequency for the oscillations of $\delta h_2$ (in cosmic time)
\be{freqRandM}
\omega_{m,r}^{\rm GR}=\frac{\M}{\sqrt{3\alpha}}
\ee
and the result is the same for both radiation and matter domination.\\
\subsection{Cosmological constant domination}
Let us finally consider the case when the cosmological constant dominates, one then has $\rho_m/\rho\ll 1$, $\rho_\Lambda/\rho\simeq 1$. Further, the dominant cosmological fluid (cosmological constant) does not evolve and it is convenient to search for the solutions in terms of the e-fold number. The Friedmann equation for the perturbation is
\be{dh2LN}
\delta h_{2,NN}+3\delta h_{2,N}+\frac{\M^4}{\alpha \rho_\Lambda}\delta h_2=0
\ee
and its general solution is
\be{solLN}
\!\delta h_{2}(N)=\frac{\bar A_1\re^{\sqrt{\frac{9}{4}-\frac{\M^4}{\alpha\rho_\Lambda}}N}\!+\bar A_2\re^{-\sqrt{\frac{9}{4}-\frac{\M^4}{\alpha\rho_\Lambda}}N}}{\re^{\frac{3N}{2}}}
\simeq \frac{A_1\sin\pa{\frac{\M^2N}{\sqrt{\alpha\rho_\Lambda}}}+A_2\cos\pa{\frac{\M^2N}{\sqrt{\alpha\rho_\Lambda}}}}{\re^{\frac{3N}{2}}}
\ee
On calculating the frequency of oscillation in terms of the cosmic time one finds the same result as was found for both matter and radiation domination, $\omega_{\Lambda}^{\rm GR}=\omega_{m,r}^{\rm GR}$.
\section{Scale Invariant Model}
Let us now consider a simple scale invariant generalisation of the action (\ref{grS}) with the addition of a scalar field $\sigma$ which dynamically generates Newton's constant. Therefore let us start from the following scale invariant action
\be{igS}
S=\frac{1}{2}\int \rd^4x\sqrt{-g}\pa{\gamma \sigma ^2R+\frac{\alpha R^2}{2}-\frac{\lambda}{2}\sigma^4-(\partial \sigma)^2 }. 
\ee
For $\alpha=0$ and in the presence of cosmological fluids (dust and radiation) the scalar field behaves as quintessence and, for certain initial conditions, it relaxes to a constant attractor and plays the role of Dark Energy. We are interested in the modification to the cosmological evolution associated with the presence of an $R^2$ contribution. \\
The independent equations for the matter gravity system are the continuity equations (\ref{Teq}) for the cosmological fluids, the Klein-Gordon (KG) equation for the homogeneous field and a ``dynamical'' Friedmann equation.
It is convenient to redefine $y\equiv \sigma^2$. In terms of $y$ the KG equation takes the form
\be{kgy}
\frac{y_{,NN}}{y}-\frac{y_{,N}^2}{2y^2}+\frac{(H^2)_{,N}}{2H^2}\frac{y_{,N}}{y}+3\frac{y_{,N}}{y}-6\gamma\frac{(H^2)_{,N}}{H^2}+2\lambda\frac{y}{H^2}-24\gamma=0
\ee
and the Friedmann equation can be rewritten as
\be{fry}
1=\frac{\rho+\frac{\lambda y^2}{4}}{3\gamma y H^2}-\frac{y_{,N}}{y}+\frac{1}{24\gamma}\pa{\frac{y_{,N}}{y}}^2-\frac{3\alpha \paq{(H^2)_{,NN}-\pa{\frac{(H^2)_{,N}^2}{2H^2}}+3(H^2)_{,N}}}{\gamma y}.
\ee
Analogously to the GR case one can write the equation for the Ricci scalar, in the IG context, which takes the following form
\be{scalIG}
\ddot R+3 H\dot R+\frac{\gamma(1+6\gamma)x}{3\alpha}R=\frac{\rho-3P+(1+6\gamma)\pa{\lambda x^2-\frac{\dot x^2}{4x}}}{3\alpha}
\ee
in terms of the cosmic time.\\
Let us now proceed following the same approach as illustrated in the previous section. We shall then consider different cosmological eras and study the departure from the evolution with $\alpha=0$.\\
\subsection{Radiation domination}
During radiation domination $\rho\simeq \rho_r$ and contributions proportional to $\lambda$ may be neglected. For $\alpha=0$ one has the solution $y=\M^2/\gamma$, $H^2\propto \re^{-4N}$ which exactly reproduces the standard $\Lambda$CDM evolution. When $\alpha\neq 0$ one expects perturbed solutions of the form
\be{pert}
H^2=\frac{\rho}{3\M^2}\pa{1+\delta h_2}, \quad y=\frac{\M^2}{\gamma}\pa{1+\delta y}
\ee
and on linearising the dynamical equations one finds
\be{linIGKGrd}
\left\{
\begin{array}{l}
6\gamma \rho^2 \delta h_{2,\rho\rho}+3\rho \delta y_{,\rho}+4\rho^2\delta y_{,\rho\rho}=0\\
4\rho^2 \delta h_{2,\rho\rho}-\frac{\M^4}{\alpha}\delta y_{,\rho}+7\rho \delta h_{2,\rho}+\frac{\M^4}{4\rho \alpha}\pa{\delta h_2+\delta y}=0
\end{array}
\right.
\ee
The first equation in (\ref{linIGKGrd}) can be solved exactly leading to
\be{sollinIGKGrd}
\delta h_2=c_1+\frac{\delta y}{6\gamma}-\frac{2\rho \delta y_{,\rho}}{3\gamma}.
\ee
The second equation in (\ref{linIGKGrd}) can be written in terms of $\delta y$ and its derivatives and has the following solution in terms of Hypergeometric functions
\ba{dxRD}
\delta y&=&-\frac{6\gamma}{1+6\gamma}c_1+c_2 \!\!\!\!\phantom{A}_1 F_2\pa{\frac{1}{4};\frac{2}{3},\frac{5}{6};-\frac{x^6}{16}}\nonumber\\
&&+c_3 x \!\!\!\!\phantom{A}_1 F_2\pa{\frac{5}{12};\frac{5}{6},\frac{7}{6};-\frac{x^6}{16}}+c_4 \!\!\!\!\phantom{A}_1 F_2\pa{\frac{7}{12};\frac{7}{6},\frac{4}{3};-\frac{x^6}{16}}
\ea
where $x=\paq{\M^4/(\alpha \rho)}^{1/6}$ and is very large for $\alpha$ small and realistic values of $\rho_r$.
In the large $x$ limit Eq. (\ref{dxRD}) takes the compact form
\be{dxRDsim}
\delta y=\bar c_2\pa{\frac{\alpha \rho}{\M^4}}^{1/4}\!\!\!+\paq{\bar c_3\sin\pa{\frac{\pi}{24}-\frac{\M^2}{\sqrt{4\alpha \rho}}}+\bar c_4\cos\pa{\frac{\pi}{24}-\frac{\M^2}{\sqrt{4\alpha \rho}}}}\pa{\frac{\alpha \rho}{\M^4}}^{3/8}\!\!\!.
\ee
and correspondingly one finds
\be{dh2expRD}
\delta h_2=\pa{\frac{\M^4}{\alpha\rho}}^{1/8}\frac{-\bar c_3 \cos\pa{\frac{\pi}{24}-\sqrt{\frac{\M^4}{4\alpha \rho}}}+\bar c_4\sin\pa{\frac{\pi}{24}-\sqrt{\frac{\M^4}{4\alpha \rho}}}}{6\gamma}
\ee
and we note that the perturbed Hubble parameter oscillates around the $\Lambda$CDM solution analogously to what occurs in (\ref{solR}). In this latter case, however, the amplitude increases much more slowly. Finally one can calculate the oscillation frequency w.r.t. cosmic time and obtain $\omega_r=\M/\sqrt{3\alpha}4$ which is the same as Eq. (\ref{freqRandM}). Let us note that, as in the model studied in Sec. II, the increasing amplitude introduced a fine tuning problem related to the initial conditions of the oscillating quantities. For this specific case the effect of the perturbative production of $\sigma$ field quanta will be addressed in the last section and can alleviate such a fine tuning issue. 
\\
\subsection{Matter domination}
During matter domination $\rho\simeq \rho_m$, and in the $\alpha\rightarrow 0$ limit one has the following solution 
\be{solmdIG0}
y=y_0 \re^{\frac{4\gamma}{1+4\gamma}N},\quad \rho=\rho_{0}\re^{-3N}, \quad H^2=\frac{1+4\gamma}{1+\frac{34}{3}\gamma+32\gamma^2}\frac{\rho}{3 \gamma y}
\ee
with $y_0\sim \M^2/\gamma$. Therefore for $\alpha$ small we expand around it:
\be{expIGm}
y=y_0 \re^{\frac{4\gamma}{1+4\gamma}N}\pa{1+\delta y},\quad H^2=\frac{(1+4\gamma)^2}{1+\frac{34}{3}\gamma+32\gamma^2}\frac{\rho\,\re^{-\frac{4\gamma}{1+4\gamma}N}}{3 \gamma y_0}\pa{1+\delta h_2}.
\ee
The linearised equations for $\delta y$ and $\delta h_2$ are
\be{kglinIGmat}
2(1+4\gamma)\delta y_{,NN}+(3+16\gamma)\delta y_{,N}-8\gamma(1+6\gamma)\delta h_{2,N}=0
\ee
\ba{frlinIGmat}
&&\delta h_{2,NN}-\frac{3(1+8\gamma)}{2(1+4\gamma)}h_{2,N}+\frac{2\M^4(3+16\gamma)(1+6\gamma)^2\re^{\frac{8\gamma}{1+4\gamma}N}}{9\alpha \rho(1+4\gamma)^3}\delta y_{,N}\nonumber\\
&&+\frac{(3+16\gamma)\paq{\frac{\M^4}{9\alpha \rho}\frac{(3+16\gamma)(1+6\gamma)^2}{(1+4\gamma)^2}\re^{\frac{8\gamma}{1+4\gamma}N}-\frac{3}{4}}}{(1+4\gamma)^2}\pa{\delta h_2+\delta y}=0.
\ea
and Eq. (\ref{kglinIGmat}) has a solution
\be{solkglinIGmat}
\delta h_2=c_1+\frac{3+16\gamma}{8\gamma(1+6\gamma)}\delta y+\frac{1+4\gamma}{4\gamma(1+6\gamma)}\delta y_{,N}
\ee
which can be substituted into Eq. (\ref{frlinIGmat}) obtaining a linear, third order, equation for $\delta y$ with the general solution
\ba{dexsolmatIG}
\!\!\!\delta y&\!\!\!=&\!\!\!-c_1\frac{8\gamma(1+6\gamma)}{3(1+4\gamma)^2}+c_2\!\!\!\phantom{A}_1F_2\pa{b;1-2b,1-b;-z} \nonumber\\
&&\!\!\!+c_3x \!\!\!\phantom{A}_1F_2\pa{2b;1-b,1+b;-z}+c_4 x^2\!\!\!\phantom{A}_1F_2\pa{3b;1+b,1+2b;-z}
\ea
where $b=\frac{4\gamma}{3(3+20\gamma)}$ and the time dependent argument of the hypergeometric functions is 
\be{zchange}
z=\frac{(3+16\gamma)(1+6\gamma)^2}{(3+20\gamma)^2} \frac{\gamma^2y^2}{3\alpha\rho}
\ee
For very large (negative) values of such an argument one observes that the hypergeometric functions can be expanded and the solution takes the compact form
\be{dexsolmatIGser2}
\delta y=\frac{\bar c_2}{z^{b}}+z^{2b-\frac{3}{4}}\paq{\bar c_3 \cos\pa{2\sqrt{z}}+\bar c_4 \sin\pa{2\sqrt{z}}}-\frac{8\gamma(1+6\gamma)c_1}{3(1+4\gamma)^2}
\ee
We observe that, apart from a constant contribution, the time dependent part decreases in time for $\gamma$ small and the oscillating part depends on the square root of the ratio between the effective (unperturbed) Planck mass given by $\gamma^2y^2$ and $\alpha \rho$ where $\rho_m$ is assumed to be the dominant fluid. This is a general feature of the solutions we found so far perturbatively and we note that the ratio is tiny leading to ``wild'' oscillations as $N$ varies. \\
On now substituting the solution for $\delta y$ in (\ref{solkglinIGmat}) and retaining the leading order contributions one finds
\be{dh2md}
\delta h_2=\bar c_0+\frac{\bar c_2\pa{9+40\gamma}}{24\gamma(1+6\gamma)z^{\frac{4\gamma}{3(3+20\gamma)}}}+\frac{(3+20\gamma)\paq{\bar c_3 \sin\pa{2\sqrt{z}}+\bar c_4 \cos\pa{2\sqrt{z}}}}{4\gamma(1+6\gamma)z^{\frac{1}{4}-\frac{8\gamma}{3(3+20\gamma)}}}
\ee
where, the leading order, non-constant, contribution decreases and is non-oscillating. Let us note that if one expresses the oscillating functions in terms of cosmic time for this case the frequency is not constant in contrast with the case of matter dominance discussed in the previous section. Indeed because of the time varying Newton's constant (to leading order) and the dependence of $z$ on the ratio $y^2/\rho_m$ (and not on $y/\rho_m\propto H^{-1}\propto t$) one has $\sqrt{z}\neq \omega_m t$. However the time dependence is tiny and proportional to $\gamma$ which must be small. On neglecting $\gamma$ one recovers the GR result i.e. $\omega_m=\omega_r+\mathcal{O}(\gamma)$ where the contribution $\mathcal{O}(\gamma)$ depends on cosmic time.\\
\subsection{Scalar field domination}
Let us finally examine the case in which the scalar field energy density dominates. In such a case the unperturbed solution ($\alpha\rightarrow 0$) mimics a cosmological constant (see \cite{FTV}) and one has
\be{s0cc}
y=\frac{\M^2}{\gamma}, \quad H^2=\frac{\lambda \M^2 }{12\gamma^2}.
\ee
On now considering $\alpha$ small the above solution is modified as follows
\be{s1cc}
y=\frac{\M^2}{\gamma}\pa{1+\delta y}, \quad H^2=\frac{\lambda \M^2 }{12\gamma^2}\pa{1+\delta h_2}.
\ee
The linearised equations for $\delta y$ and $\delta h_2$ are
\be{frcceq}
\left\{\begin{array}{l}
\delta h_{2,NN}+3 \delta h_{2,N}+\frac{4\gamma^2}{\alpha \lambda}\pa{\delta y_N-\delta y+\delta h_2}=0\\
\\
\delta y_{NN}+3 \delta y_{,N}+6\gamma\pa{4\delta y-\delta y_{2,N}-4\delta y_2}=0
\end{array}
\right.
\ee
which can be solved exactly. The general solutions are 
\be{solcc1}
\left\{
\begin{array}{l}
\delta h_2=A_2+A_1\re^{-3N}+\re^{-\frac{3}{2}N}\left(A_3\re^{\frac{\Lambda}{2\alpha \lambda}N}+A_4\re^{-\frac{\Lambda}{2\alpha \lambda}N}\right)\\
\delta y=\frac{A_1}{4}\re^{-3N}+A_2+\frac{\pa{3\Lambda-15\alpha \gamma}}{4\gamma(1+6\gamma)}\re^{-\frac{3}{2}N}\pa{A_3\re^{\frac{\Lambda}{2\alpha \lambda}N}+A_4\re^{-\frac{\Lambda}{2\alpha \lambda}N}}
\end{array}
\right.
\ee
where 
\be{Lambda}
\Lambda\equiv \sqrt{-\alpha\lambda\pa{16\gamma^2+96\gamma^3-9\alpha\lambda+96\alpha\gamma\lambda}}.
\ee
For realistic values of the parameters and taking $\alpha\lambda\ll\gamma\ll1$, $\Lambda$ is pure imaginary and approximately given by $\Lambda\simeq 4\gamma\sqrt{\alpha\lambda}\,i$.
Therefore after few e-folds $\delta h_2$ and $\delta y$ are oscillating functions multiplied by a damping factor
\be{deltah2appcc}
\left\{
\begin{array}{l}
\delta h_2\sim \re^{-\frac{3}{2}N}\left[\bar A_3\cos\pa{\frac{2\gamma}{\sqrt{\alpha \lambda}}N}+\bar A_4\sin\pa{-\frac{2\gamma}{\sqrt{\alpha \lambda}}N}\right]\\
\delta y\sim -\frac{15\alpha }{4(1+6\gamma)}\re^{-\frac{3}{2}N}\left[\bar A_3\cos\pa{\frac{2\gamma}{\sqrt{\alpha \lambda}}N}+\bar A_4\sin\pa{-\frac{2\gamma}{\sqrt{\alpha \lambda}}N}\right]
\end{array}
\right. .
\ee
In terms of the cosmic time $N\simeq \sqrt{\frac{\lambda}{3}}\frac{\M}{2\gamma}(t-t_0)$, the oscillation frequency is the same transplanckian frequency as one finds in GR: $\omega_{\Lambda}=\M/\sqrt{3\alpha}$.
\section{Perturbative Particle Production}
The highly oscillatory behaviour of the homogeneous dynamical system will lead to particle production similarly to that occurring at the end of inflation, during the so-called reheating era \cite{reheating}. In this latter case the oscillatory behaviour of the inflaton condensate makes it decay into SM particles (or more exotic stuff) which ends up reheating the Universe if the inflaton couples to such particles. Gravity interacts with every particle and, for the case we are considering such a coupling certainly exists. The particle production due to an oscillating Ricci scalar in $f(R)$ gravity models has been studied in several articles \cite{Dolgov}. Here we shall limit ourselves to the study of the production of scalar particles associated with the field $\sigma$. The field is non-minimally coupled and $\sigma$ field quanta production should be non-negligible, at least for large enough values of the coupling $\gamma$. A significant production of such quanta may modify cosmological evolution and dampen the oscillations of the homogenous sector, thus it is necessary to, at least, provide an estimate of such effects. Since the amplitude of the oscillations of the homogeneous sector is increasing only in a radiation dominated universe we shall restrict our analysis to that era. At later times, after radiation domination, oscillations are small, non-increasing and are quickly dampened by particle production. The effects of particle production may thus be neglected.\\
The equation for $\sigma(t,\vec x)$ is the KG equation
\be{fulleqs}
\Box \sigma-\gamma R\sigma+\lambda \sigma^3=0
\ee
where $\Box\equiv g_{\mu\nu}\nabla^\mu\nabla^\nu$. On now defining $\sigma(t,\vec x)=\sqrt{y(t)}+\delta \sigma(t,\vec x)$, where $y(t)$ satisfies the homogeneous equations, one is led the following linearised equation (we neglect metric perturbations):
\be{perteq}
\Box \delta \sigma-6\gamma \pa{2H^2+\dot H}\delta \sigma+3\lambda y\,\delta \sigma=0.
\ee
and $H^2$ and $y$ are given by (\ref{pert}) during radiation domination, by (\ref{solmdIG0}) during matter domination and by (\ref{s1cc}) during scalar field domination. Thus 
\be{dH}
\dot H=\frac{1}{2}\pa{H^2}_{,N}=\frac{1}{2}\paq{\frac{\rho_{,N}}{\rho}-\frac{y_{0,N}}{y_0}+\frac{\pa{\delta h_2}_{,N}}{1+\delta h_2}}H^2.
\ee
where $y_0$ here denotes the unperturbed (LO) quantity and $y_{,N}/y=0$ during radiation and scalar field domination and $y_{,N}/y=4\gamma/(1+4\gamma)$ during matter domination.\\
On now inserting Eq. (\ref{dH}) in Eq. (\ref{perteq}) and retaining contributions to first order in $\delta h_2$ and $\delta y$ one obtains
\be{perteq2}
\Box \delta \sigma-3\gamma \paq{4-3(1+w)-\frac{\dot y}{Hy}}H^2(t)\delta \sigma+3\gamma\,\dot{\delta h_2}\,H_0(t)\delta \sigma+3\lambda y\,\delta \sigma=0.
\ee
where $w$ is the equation of state of the fluid which drives the universe expansion and $H_0(t)$ is the leading order time dependence in $H(t)$ (i.e. the Hubble parameter when $\alpha=0$). During radiation dominance $H_0(t)$ is given by the standard, $\Lambda$CDM expression $H_0=\rho_r/(3\M^2)$ but during matter dominance it is given by the cumbersome (and non-standard) expressions in Eq.(\ref{solmdIG0}). Let us note that the first bracket of Eq. (\ref{perteq2}) cancels in a radiation dominated universe. Essentially Eq. (\ref{perteq2}) is an equation for a scalar field with a (slowly) time varying ``effective'' mass and an interaction with an ``external'' oscillating field. The interaction may lead to a significant production of quanta (particles) of the field $\delta \sigma$. To the leading order such particles are produced in pairs. Let us note that the typical time scale of the oscillations is much smaller than that associated with the variation of the ``effective'' mass which, in our rough estimates, will be taken as constant. Moreover since the oscillation is much faster than the expansion rate of the universe the decay rate can be calculated in the flat space limit. This latter assumption is generally valid also when one calculates the decay rate of the inflaton at the end of inflation. More refined treatments can be found in the literature, taking into account, for example, the tiny higher order effects which lead to the distortion of the harmonic solution, or its time dependence (which here is neglected). Essentially such diverse approaches give more refined results but coincide in the LO with ours.
\subsection{Particle Production from an oscillating field}
Let us briefly review how to estimate the particle production due to the presence of a coupled, oscillating field and consider, for simplicity, just the case of two interacting scalar fields. Let us suppose the interaction potential between such fields has the form $V=g \,m\, \chi^2\,\phi$ where $m$ is some mass scale and $\phi(t)$ is an external, homogeneous, scalar field with $\phi(t)=\bar \phi \cos\pa{\omega t}$. In flat space time the Klein-Gordon equation for $\chi$ is
\be{chieq}
\Box \chi+\mu^2\chi+2g\, m\, \chi\,\phi=0.
\ee
If one calculates the transition amplitude (in the interaction picture), then to the lowest order the amplitude for $\phi$ to decay into two $\chi$ particles is
\be{tramp}
{\mathcal A}_{\phi\rightarrow \chi\chi}\equiv \langle \vec k_1,\vec k_2|\re^{-i\int\rd x^4 V_{\rm int}}|0\rangle\simeq -i (g\,m)\int \rd t\phi(t)\int\rd^3 x\langle \vec k_1,\vec k_2|\chi^2|0\rangle.
\ee
On now expanding $\chi$ in terms of creation-annihilation operators:
\be{fouchi}
\chi=\int \frac{\rd^3 k}{(2\pi)^{3/2}\sqrt{2\omega_k}}\pa{\hat a_{\vec k}\re^{-i\omega_k t+i\vec k\cdot \vec x}+{\rm h.c.}},
\ee
with $\omega_k^2=\vec k^2+m^2$, and integrating over space ($\vec x$) and time one finds
\ba{tramp2}
{\mathcal A}_{\phi\rightarrow \chi\chi}&\simeq& -\frac{2\pi i (g\,m)}{\omega_{k_1}} \delta^{(3)}\pa{\vec k_1+\vec k_2}\int\rd t\frac{1}{2\pi}\frac{\bar\phi \pa{\re^{i\omega t}+\re^{-i\omega t}}}{2}\re^{2i\omega_{k_1}t}\nonumber\\
&\simeq& -\frac{\pi i (g\,m)\bar\phi}{\omega} \delta^{(3)}\pa{\vec k_1+\vec k_2} \delta{\pa{\omega_{k_1}-\frac{\omega}{2}}}.
\ea

The probability can then be calculated from the modulus squared of (\ref{tramp2})
\be{prob1}
\left|{\mathcal A}_{\phi\rightarrow \chi\chi}\right|^2=(2\pi)^4\delta^{(4)}(0)\frac{ (g\,m)^2\bar\phi^2}{16\pi^2\omega^2}\delta^{(3)}\pa{\vec k_1+\vec k_2} \delta{\pa{\omega_{k_1}-\frac{\omega}{2}}}
\ee
where, as usual, one interprets the divergent $\delta^{(4)}(0)$ as the integral over all spacetime $(2\pi)^4\delta^{(4)}(0)=VT$.
On now dividing by $VT$ one obtains the transition probability per unit time and volume:
\be{dP}
\rd P(\vec k_1,\vec k_2)=\frac{ (g\,m)^2\bar\phi^2}{16\pi^2\omega^2}\delta^{(3)}\pa{\vec k_1+\vec k_2} \delta{\pa{\omega_{k_1}-\frac{\omega}{2}}}.
\ee
The decay rate of the oscillating field quanta can now be calculated using energy conservation: from each $\phi$ particle decay one obtains two $\chi$ particles with the same energy $\omega/2$. The number of $\chi$ particles produced with momenta $\vec k_1$ and $\vec k_2$ is proportional to Eq. (\ref{dP}) and the energy lost by the oscillating scalar field and transferred to the $\chi$ field is also proportional to the same quantity. In particular, the energy variation per unit time of the fluid of $\chi$ particles due to the decay of a quantum with energy $\omega$ is given by
\be{dE/dt}
\frac{\rd E_\chi}{\rd t}=\omega\,V\,\rd P=V\,\frac{ (g\,m)^2\bar\phi^2}{16\pi^2\omega}\delta^{(3)}\pa{\vec k_1+\vec k_2} \delta{\pa{\omega_{k_1}-\frac{\omega}{2}}}.
\ee
The total variation in time of the energy density $\dot\rho_\chi=\dot E_\chi/V$ is obtained by integrating the last expression over the final momenta and thus
\ba{drho/dt}
\dot \rho_\chi=-\dot \rho_\phi&=&\int\rd^3k_1\rd^3k_2 \frac{ (g\,m)^2\bar\phi^2}{16\pi^2\omega}\delta^{(3)}\pa{\vec k_1+\vec k_2} \delta{\pa{\omega_{k_1}-\frac{\omega}{2}}}\nonumber\\
&=&\sqrt{1-\frac{4\mu^2}{\omega^2}} \frac{ (g\,m)^2\bar\phi^2\omega}{16\pi}.
\ea
The energy density associated with the oscillating field $\phi$, to the LO in $g$, is $\rho_\phi=\omega^2\bar\phi^2/2$
and the decay rate is finally
\be{decayrate}
\Gamma_{\phi\rightarrow \chi\,\chi}=-\frac{\dot\rho_\phi}{\rho_\phi}=-\sqrt{1-\frac{4\mu^2}{\omega^2}} \frac{ (g\,m)^2}{8\pi\omega}.
\ee
If one now takes into account the expansion of the Universe then the amplitude of the oscillating field varies in time with a rate typically much slower that the frequency of the oscillating part. In such a case $\bar \phi\rightarrow \bar \phi(t)$. The decay rate, however, is independent of $\bar \phi$ and the same holds for an expanding Universe. Therefore in such an approximation $\Gamma_{\phi\rightarrow \chi\,\chi}$ is that found on flat space. The presence of a coupling between an oscillating function and the field $\delta \sigma$ leads to perturbative particle production with transplanckian energy. Such particles may ``re-heat'' the Universe significantly by adding radiation like matter to the total energy content. This mechanism has been discussed already in the literature (see \cite{Dolgov}) but not in our context, where, in particular, the non-minimal coupling between ``matter'' (the field $\sigma$) and gravity is a feature of the model and is not simply guessed at or arbitrarily introduced.

\subsection{Radiation Domination}
We now estimate the production rate and its back-reaction in a radiation dominated universe. In such a case several simplifications occur in the perturbation equation (\ref{perteq2}) which takes the form:
\be{perteqRD}
\ddot{\delta \sigma}+3H(t)\dot{\delta \sigma}-\frac{\nabla^2}{a^2} \delta \sigma+3\gamma \frac{\dot y}{y}H(t)\delta \sigma+3\gamma\,\dot{\delta h_2}\,H_0(t)\delta \sigma+3\lambda y\,\delta \sigma=0.
\ee
where
\be{Hrd}
\!\!\!\!\!H(t)\simeq H_0(t)\pa{1+\frac{\delta h_2}{2}}\;{\rm with}\; H_0(t)\simeq \sqrt{\frac{\rho_r}{3\M^2}}\;{\rm and}\; y\simeq \frac{\M^2}{\gamma}\pa{1+\delta y}.
\ee
Let us note that $\delta h_2$ and $\delta y$ are oscillating functions given by Eqs. (\ref{dxRDsim},\ref{dh2expRD}). The frequency of oscillation can be easily calculated on observing that the arguments of the trigonometric functions in such expressions are inverse proportional to the square root of $\rho_r$. The frequency of oscillation is (\ref{freqRandM}) and for $\alpha$ small is transplanckian. Moreover we note that the scale factor also is an oscillating function which, in principle can be obtained on integrating the Hubble constant. Thus several terms in (\ref{perteqRD}) have an oscillatory behaviour and may lead to particle production. To the leading order, the $\delta \sigma$ field quanta are produced in pairs with transplanckian energies $E_{\delta\sigma}=\omega_r/2$ and may thus contribute to the total radiation content. If we just consider the interaction terms of the form $A(t)\cos (\omega_r t)\delta \sigma$ in Eq. (\ref{perteqRD}), we then observe that 
\be{incrpert}
3\gamma\,\dot{\delta h_2}\,H_0(t)\delta \sigma\sim \frac{\bar c_4}{2}H_0(t)^2\pa{\frac{\M^4}{\alpha \rho}}^{5/8} \cos \pa{\omega_r t}\,\delta \sigma
\ee
is the contribution which has the slowest rate of decrease in an expanding universe\footnote{Observe that $\dot\delta h_2$ increases but $H_0^2$ decreases and the total amplitude is decreasing.}. On comparing it with Eq. (\ref{chieq}), from Eq. (\ref{drho/dt}) we have
\be{drhosigna}
\dot \rho_{\delta \sigma}\simeq \frac{\omega_r}{256\pi} \bar c_4^2H_0(t)^4\pa{\frac{\M^4}{\alpha \rho}}^{10/8}= \frac{\bar c_4^2}{256\pi}\pa{\frac{\M}{\sqrt{3\alpha}}}^5\pa{\frac{\alpha\rho_r}{\M^4}}^{3/4}
\ee
where $\bar c_4$ is an integration constant. The energy density added by the perturbative production of the $\delta \sigma$ quanta contributes to the total radiation content in the Universe and the continuity equation for radiation must be modified accordingly into
\be{rhoRAD}
\rho_{r}\rightarrow\rho_r+\rho_{\delta \sigma}\Rightarrow \dot \rho_{r}\equiv -4H_0\rho_r+\frac{\bar c_4^2}{256\pi}\pa{\frac{\M}{\sqrt{3\alpha}}}^5\pa{\frac{\alpha\rho_r}{\M^4}}^{3/4}.
\ee
Correspondingly the amplitude (and energy density) of the ``driving fields'' $\delta h_2$ and $\delta y$ decays faster than that previously obtained without including the effect of particle production. In principle the modified decay rate can be estimated by energy conservation. 
\begin{figure}[t!]
\centering
\epsfig{file=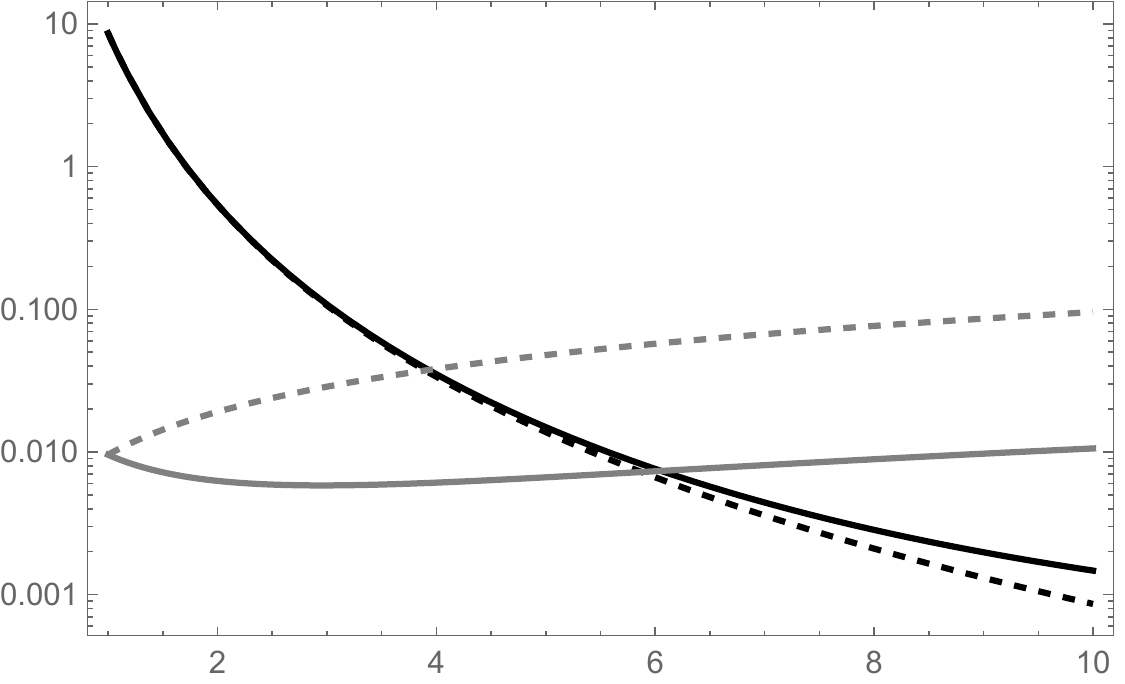, width=8.5 cm}
\caption{{\it The figure shows the evolution of $f_r(a)$ (black) and $f_{\delta h_2}(a)$ (gray) with$A_0=\re$ and $B_0=10^{-3}$. The solid lines are the solutions for $C_0=10^{-2}$ and the dashed lines are the unperturbed solutions ($C_0=0$). The radiation fluid decreases slower than the usual $a^{-4}$ behaviour and the energy density of $\delta h_2$ increases very slowly.} 
\label{solf}}
\end{figure}
Let us therefore estimate the back-reaction of particle production on the energy density associated to the oscillating homogeneous field $\delta h_2$.
We first observe that the energy density associated with the oscillation of $\delta h_2$ can be estimated in analogy with that of the oscillating inflaton during reheating. In this latter case it is proportional to its mass squared and its amplitude squared (apart from dimensional factors). The mass of $\delta h_2$ is its frequency ($\omega_r$) and thus $\rho_{\delta h_2}\sim A_{\delta h_2}^2\omega_r^2/2$.
From energy conservation the continuity equation for $\rho_{\delta h_2}$ is
\be{condsigma}
 \dot \rho_{\delta h_2}=2\frac{\dot A_{\delta h_2}}{A_{\delta h_2}}\rho_{\delta h_2}-\dot \rho_{\delta \sigma}
\ee
where the ratio $2\dot A_{\delta h_2}/A_{\delta h_2}\equiv \Gamma_{\delta h_2}^{(0)}$ is that calculated without taking into account the perturbative decay. On now using the solution (\ref{dh2expRD}) one has
\be{dAh2}
A_{\delta h_2}\sim\pa{\frac{\M^4}{\alpha \rho}}^{1/8}\Rightarrow \Gamma_{\delta h_2}^{(0)}=H_0=\sqrt{\frac{\rho_r}{3\M^2}}.
\ee
One is then led to the continuity equation
\be{decaydy}
\dot \rho_{\delta h_2}=-H_0\rho_{\delta h_2}-\frac{\bar c_4^2}{256\pi}\pa{\frac{\M}{\sqrt{3\alpha}}}^5\pa{\frac{\alpha\rho_r}{\M^4}}^{3/4}
\ee
and on solving the system of coupled equations (\ref{rhoRAD}, \ref{decaydy}) one obtains the evolution of the radiation fluid and that of the time dependent amplitude $A_{\delta h_2}^{\rm pert}$. In terms of the scale factor one has the following system
\be{sys}
\left\{
\begin{array}{l}
a\frac{\rd f_r}{\rd a}=-4 f_r+R_0 f_r^{1/4}\\
\\
a\frac{\rd f_{\delta h_2}}{\rd a}= f_{\delta h_2}-R_0 f_r^{1/4}
\end{array}
\right.
\ee
where $f_r=\rho_r/\M^4$ and $f_{\delta h_2}=\rho_{\delta h_2}/\M^4$ and $R_0$ is a constant which depends on the integration constant $\bar c_4$. The system has the following general solution 
\be{solsys}
f_r=\pa{\frac{A_0^3+C_0 a^3}{4a^3}}^{4/3},\;\,f_{\delta h_2}=b_0 a+\frac{C_0\pa{C_0+\frac{A_0^3}{a^3}}^{1/3}\!\!\!\!\!\!\!\!\phantom{A}_2 F_1\pa{-\frac{2}{3},-\frac{1}{3};\frac{1}{3};-\frac{C_0 a^3}{A^3}}}{2^{5/3}\pa{1+C_0 a^3/A_0^3}}
\ee
where $A_0$ and $B_0$ are integration constants. The general solutions Eq. (\ref{solsys}) are illustrated in the Figure (\ref{solf}), for a particular choice of the model parameters, and compared with the solutions without particle production. At $a=1$ the initial conditions are such that the solutions coincide and the figure plots their difference as time evolves. The black line describes the energy density of radiation in Planck mass units with and without particle production (solid and dashed lines respectively) while the grey lines plot the energy density of $\delta h_2$ in Planck mass units with and without particle production (solid and dashed lines respectively). It is evident that particle production increases the radiation energy density which differs from the usual $a^{-4}$ scaling behaviour, and lowers the energy density of $\delta h_2$. Since the frequency of $\delta h_2$ remains constant, such a decrease just affects the amplitude of its oscillations. Let us note that the decay rate decreases as the universe expands and its effects becomes less and less relevant as time evolves. The injection of very energetic particles in the plasma may also have consequences on Nucleosynthesis. The $\sigma$ field, however, is coupled to SM particles indirectly and its effects might be small enough. Such a point certainly deserves further examination and we leave it to future studies. 
\section{Asymptotic Behaviour}
In this last section we study the different asymptotic behaviour of the two models (on neglecting any particles production effect which are shown to be negligible for $a$ large). This analysis is performed by searching for the fixed points of the two autonomous systems with different definitions of the dimensionless dynamical variables. Starting from the perturbed $\Lambda$CDM case, let us rewrite the Friedmann-like equation as follows
\be{frGR2}
1-x_1-x_2-x_3+x_4-\frac{\M^4}{\alpha\rho_\Lambda}x_1x_4^2+\frac{\alpha R_{,N}}{\M^2}=0.
\ee
where we defined
\be{def1x}
x_1=\frac{\rho_\Lambda}{3H^2\M^2},\;x_2=\frac{\rho_m}{3H^2\M^2}\;x_3=\frac{\rho_r}{3H^2\M^2};\;x_4=\frac{\alpha R}{\M^2}
\ee
We can differentiate w.r.t. $N$ the dimensionless $x_i$'s and use the dynamical equations (\ref{Teq},\ref{frGR2}). Then, on setting such derivatives to zero, one obtains the fixed points of the system. The following solutions are found
\be{solGR1}
x_1=x_2=0,\,x_3=1+x_4\quad {\rm and} \quad x_1=1,\, x_2=x_3=0,\,x_4=\frac{4\alpha \rho_L}{\M^4}
\ee
and the first solution is $3\M^2H^2=\rho_{r}$ and exists if $\rho_\Lambda=\rho_M=0$. The second solution is the DE attractor in the future with $3\M^2H^2=\rho_\Lambda$. These solutions have already been discussed in Sec. II during radiation domination and Dark Energy domination and also exist for the ``unperturbed'' system.\\
On instead defining $x_4=9\alpha (H^2)_{,N}/\M^2$ and following the same procedure as illustrated above, one also obtains
the solution
\be{solGR21}
H^2=-\frac{\M^2}{9\alpha} N+h_2
\ee
which is peculiar to the perturbed $\Lambda$CDM system we are considering.\\
\subsection{Scale invariant cosmology}
For the scale invariant model the relevant dynamical equations (apart from the continuity equations for matter and radiation) are the Friedmann-like equation and the KG equation
\be{FRIG}
\left\{
\begin{array}{l}
1=\frac{\lambda y}{12 \gamma H^2}+\frac{\rho_r}{3\gamma H^2 y}+\frac{\rho_m}{3\gamma H^2 y}-\frac{\alpha R}{\gamma y}+\frac{\alpha R^2}{12\gamma H^2 y}-\frac{y_{,N}}{y}+\frac{y_{,N}^2}{24\gamma y^2}-\frac{\alpha R_{,N}}{\gamma y}\\
\frac{y_{,NN}}{24\gamma y}-\frac{y_{,N}^2}{48\gamma y^2}+\frac{y_{,N}}{24\gamma y}+\frac{R y_{,N}}{144\gamma H^2 y}+\frac{\lambda y}{12\gamma H^2}-\frac{R}{12 H^2}=0
\end{array}
\right. .
\ee
On defining the following set of dynamical variables
\be{defxiIG}
x_1=\frac{R}{y},\, x_2=\frac{R}{H^2},\, x_3=\frac{y_{,N}}{y},\, x_4=\frac{\rho_r}{3\gamma y H^2},\, x_5=\frac{\rho_m}{3\gamma y H^2} 
\ee
one obtains the following results. The first two solutions are the radiation and cosmological constant domination with $y=y_0$ and $3\gamma y_0H^2=\rho_r$ or $12\gamma H^2=\lambda y_0$ respectively, which also exist in the limit $\alpha=0$. For $\rho_r=\rho_m=0$ one obtains
\be{sol13IG}
\gamma=-\frac{1}{6},\,y=y_0\re^{-2N}\, H^2=h_2 \re^{-4N}.
\ee
and the more cumbersome solutions 
\be{sol145IG2y}
y=y_0\re^{-\frac{3+12(1-3\gamma)\gamma-36\alpha \lambda\pm\widetilde\Gamma}{2(1+6\gamma)}N},\; H^2=\frac{y_0\pa{1+6\gamma}^2\re^{-\frac{3+12(1-3\gamma)\gamma-36\alpha \lambda\pm{\widetilde\Gamma}}{2(1+6\gamma)}N}}{9\alpha\paq{5+36\pa{\gamma+\gamma^2+\alpha \lambda}\mp\widetilde\Gamma}}
\ee
where $\widetilde \Gamma=\sqrt{3\paq{3+4\gamma(5+3\gamma)+12\alpha \lambda}\paq{(1+6\gamma)^2+36\alpha\lambda}}$. Finally the solution
\be{sol16IG}
y=y_0\re^{-2N},\, H^2=\frac{\lambda y_0\re^{-2N}}{1+6\gamma},\,\alpha=-\frac{(1+6\gamma)^2\pa{3+\frac{4\rho_{r,0}}{\lambda y_0^2}}}{108\lambda}
\ee
is found for $\rho_m=0$ and the solution 
\be{sol17IG}
y=y_0\re^{-\frac{3}{2}N},\, H^2=\frac{8\lambda y_0\re^{-\frac{3}{2}N}}{3\pa{3+20\gamma}},\,\alpha=-\frac{(3+20\gamma)^2\pa{\frac{3(1+6\gamma)}{3+20\gamma}+\frac{2\rho_{m,0}}{\lambda y_0^2}}}{360\lambda}
\ee
exists when $\rho_r=0$. Let us note that the last two solutions have $\alpha$ fixed and negative. \\
On changing the definition of the dimensionless dynamical variables and setting $x_2=\frac{\lambda y}{12\gamma H^2}$ one finds two new solutions for $y=y_0$ and
\be{sol21IG}
H^2=h_2\re^{6N}-\frac{\gamma y_0}{12\alpha}\,\;{\rm or}\,\;H^2=h_2\re^{-4N}-\frac{\gamma y_0}{12\alpha}
\ee
and the solution
\be{sol22IG}
y=y_0\re^{-2N},\,H^2=h_2\re^{-4N}-\frac{(1+6\gamma) y_0}{36\alpha}\re^{-2N}
\ee
valid when $\lambda=\rho_m=\rho_r=0$. A third, possible, definition of the dynamical variables is finally considered i.e. $x_1=\lambda y/(12\gamma H^2)$ and $x_2=9\alpha \pa{H^2}_{,N}/(\gamma y)$. In such a case, when $\lambda=\rho_m=\rho_r=0$, one finds the solution
\be{sol312IG}
y=y_0\re^{-\pa{3\pm\sqrt{9+48\gamma}}N},\,H^2=h_2+y_0\frac{3+12\gamma\pm\sqrt{9+48\gamma}}{36\alpha\pa{3\pm\sqrt{9+48\gamma}}}\re^{-\pa{3\pm\sqrt{9+48\gamma}}N}.
\ee
Given the presence of the 2 more degree of freedom associated with the field $\sigma$ and its derivative, the scale invariant model has more solution compared with the perturbed $\Lambda$CDM model. In particular many of them are realised when the cosmological fluids are absent or subdominant. 

\section{Conclusions}
We studied the cosmological evolution of a scale invariant model having a quintessence scalar field non-minimally coupled to gravity and an $R^2$ term. The scalar field dynamically generates both Newton's constant and the cosmological constant. The model is the simplest, scale invariant, generalisation to a previously analysed dark energy model \cite{FTV}. While the non-minimal coupling of the scalar field naturally arises when one considers the quantum corrections of a scalar field on a curved background, the $R^2$ contribution is generally introduced to describe quantum gravitational effects. This model has been already studied in the context of inflation \cite{Rinaldi} where the scalar field plays the role of the inflaton and we analyse here some of its dynamical features at lower energies, after the Hot Big Bang, as a model of quintessence. The $\alpha R^2$ contribution alters significantly the cosmological evolution even for tiny values of $\alpha$, adding a new scalar degree of freedom. Indeed the $00$ Einstein equation becomes a dynamical equation for $R$ and, in principle, independently of the value of $\alpha$, new cosmological solutions exist. We limit our analysis to solutions which can be obtained on slightly deforming those found in the $\alpha=0$ limit, which were already found to be viable as Dark Energy models (or at least not too different from $\Lambda$CDM) provided $\gamma\ll 1$ \cite{FTV}.\\ 
In Sec. II the cosmological evolution of $\Lambda$CDM with the addition of the ``quantum corrections'' $\alpha R^2$ is studied by perturbing the $\Lambda$CDM model ($\alpha\ll 1$). Analytical solutions are found during radiation, matter and cosmological constant domination by expanding the dynamical equation to the first order in perturbations. Such solutions share the common feature of having an oscillating contribution with a frequency $\omega=\M/\sqrt{3\alpha}$ (transplanckian). However only during radiation domination does the oscillating contribution have an increasing amplitude which may significantly alter cosmological evolution and have consequences which are not in agreement with observations. During matter domination the amplitude is constant and at later times, when the cosmological constant begins dominating, the amplitude is decreasing. As shown in \cite{Dolgov} the oscillatory behaviour of the solutions leads to perturbative particle production analogously to what occurs at the end on inflation, when the homogeneous inflaton transfers its energy to matter fields and reheats the Universe. Thus such an energy transfer damps the oscillatory solution which finally tends to the unperturbed ones, but with additional radiation injected into the system.\\
We then considered the full, scale invariant system and compared it with the former matter gravity system. On perturbing the solutions already found for $\alpha=0$, we obtained the analytical solutions to the first order in the perturbations on assuming radiation, matter and cosmological constant domination separately. If one does not consider them separately the equations for the perturbed cosmological evolution are much more cumbersome. The solutions share some relevant features with those previously obtained for the perturbed $\Lambda$CDM model. Indeed the perturbations oscillate with the same frequency during radiation and scalar field domination and the amplitude of such oscillations increases in time for the first case and decreases in the second. We observe, however, that the rate of increase during radiation domination is slower for the scale invariant model. During matter domination the evolution slightly changes since the ``dynamical'' Newton's constant varies to LO. Thus the frequency becomes slightly time dependent and the amplitude, compared with the perturbed $\Lambda$CDM model, is not constant but decreases. Finally we evaluated the effects of perturbative scalar particle production during radiation dominance. Such an analysis was already performed in the literature for the perturbed $\Lambda$CDM model in \cite{Dolgov} by considering a minimally coupled scalar field. In the scale invariant case one studies the particle production associated with the non-minimally coupled scalar field playing the role of Dark Energy and the back-reaction of such a production on the amplitude of the oscillating quantities. We just considered the LO effects of particle production on the oscillating perturbed Hubble parameter. In such a case the modified continuity equations are estimated and solved analytically. The solution shows that due to energy transfer the radiation fluid energy density has a non-standard scaling behaviour for a given time interval and correspondingly the amplitude of oscillations is damped. An efficient damping would alleviate the problem of the fine tuning of the initial conditions of the model which, otherwise, could sensibly deviate from standard cosmology and then become non-viable. A more refined analysis which takes into consideration higher order contributions to the oscillatory behaviour of the homogeneous solutions and a more complete study of the perturbative production of particle during the entire universe evolution is necessary and is left for future studies.\\
Finally, in the last section, we obtained some cosmological solutions, which exist for particular choices of the parameters and the universe content, for the two models being compared. This last part certainly deserves further in-depth analysis and a study of the stability of the solutions found should be also addressed.  

\end{document}